# Transaction Remote Release (TRR): A New Anonymization Technology for Bitcoin

QingChun ShenTu[1*], JianPing Yu [1]

[1] ATR Defense Science & Technology Lab., Shenzhen University, Shenzhen, China
[*] unshadowster@gmail.com

**Abstract:** The anonymity of the Bitcoin system has some shortcomings. Analysis of Transaction Chain (ATC) and Analysis of Bitcoin Protocol and Network (ABPN) are two important methods of deanonymizing bitcoin transactions. Nowadays, there are some anonymization methods to combat ATC but there has been little research into ways to counter ABPN. This paper proposes a new anonymization technology called Transaction Remote Release (TRR). Inspired by The Onion Router (TOR), TRR is able to render several typical attacking methods of ABPN ineffective. Furthermore, the performance of encryption and decryption of TRR is good and the growth rate of the cipher is very limited. Hence, TRR is suited for practical applications.

## 1. Introduction

Bitcoin is a decentralized crypto-currency, introduced by Satoshi Nakamoto [1] in 2008 and deployed in January 2009. Bitcoin has several characteristics including reaching most nodes' consensus by p2p protocol, decentralised production of Bitcoins by proof of work (PoW), preventing double spending by transparent transactions, pseudo-anonymity, and protecting personal privacy, which have made Bitcoin increasingly popular. The total value of Bitcoins in circulation reached a peak of $12 billion in 2013.

Research on Bitcoin anonymity originated from the idea of e-Cash. David Chaum [2] first proposed to implement an anonymous e-cash based on the use of 'blind signatures' in 1982, which is intended to protect the anonymity of the sender unconditionally. In 1992, Tatsuaki [3] proposed 6 basic characteristics of ideal e-cash, among them, privacy is an important feature.

In the Bitcoin network, Bitcoin addresses act as the user's account. Generally speaking, the aim of anonymization is to prevent attackers from discovering the relationship between Bitcoin addresses and the real or virtual user identity information through the Bitcoin network and the blockchain. Conversely, deanonymization is the uncovering of the relationship between the Bitcoin address and user in which the IP address is important user identity information.

The current study on Bitcoin deanonymization [4-12] focuses on two methods. One is the Analysis of the Transaction Chain (ATC), which is to obtain transaction information from public blockchain data, to classify Bitcoin addresses based on the weakness of Bitcoin anonymity [4-7], and to relate Bitcoin addresses to other personal identities [6-7]. The other method is the Analysis of the Bitcoin Protocol and Network (ABPN). This makes use of the spreading characteristics of Bitcoin transactions to deduce the source IP address of a new transaction [8-12].

There are several methods to combat ATC attacks such as Coinjoin [13], Ring signature and Stealth address [14], ZeroCoin [15] and ZeroCash [16]. However, apart from The Onion Router (TOR) [17], there is no way to deal with ABPN attacks. Furthermore, the Bitcoin system running over TOR has been proven to be unsafe [12]. This paper proposes a new anonymization technology: Transaction Remote Release (TRR), which can resist known attacks against Bitcoin protocols and network, including the Sybil attack [9-11] and fake Bitcoin nodes [12].

## 2. The Bitcoin System

The understanding of Bitcoin anonymity requires some basic knowledge of the Bitcoin system such as the transaction chain, Bitcoin network and protocol, blockchain and mining. Here we introduce some of the basic components of Bitcoin.

*2.1. Transactions and the transaction chain*

The Bitcoin address is an account on the Bitcoin network, which corresponds to a bank account in conventional currency systems. Bitcoin security is based on public key encryption. A Bitcoin address is generated by double hashing the public key. Only the user who owns the corresponding private key can make use of Bitcoins lodged at this address.

The Bitcoin system has a public ledger in which transferred records are stored rather than the balance of every Bitcoin address. A transferred record is known as a Bitcoin transaction. It includes the transfer time, inputs, outputs, amount, and signatures. Bitcoin transactions with their related inputs and outputs enter into the transaction chain, as shown in Fig. 1.

Every transaction has a unique id. Each input is connected with the output of the previous transaction so we can find the input address of the transaction through the output address of former one. Inputs and outputs within a transaction may be many to one, or one to many. All inputs should not be used within all existing transactions as this will prevent the successful verification of the transaction. Signatures aim to prove the input amount belongs to the sender because only the private key owner can sign the transaction properly. Other nodes verify signatures using the public key of the sender.

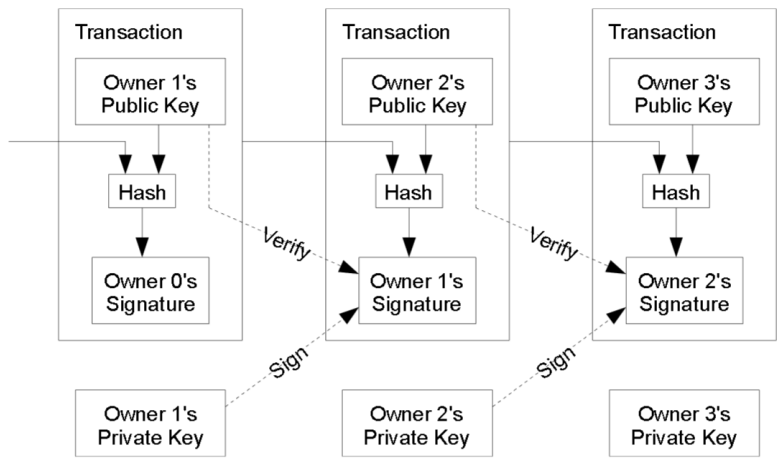

*Fig.1*. *Transactions and the chain of transactions*

## 2.2. Network and protocol

Bitcoin nodes communicate with each other via unencrypted TCP connections using port 8333. A Bitcoin wallet which does not accept incoming connections is known as the Client while others are called Bitcoin Nodes. Both the client and node save the copy of the IP addresses and communication ports of other clients and nodes. By default, they always keep 8 outgoing connections, and the node allows 117 incoming connections making 125 connections in total. If the number of outgoing connections falls below 8 they will reconnect until then number returns back to 8 entry nodes.

***Penalty points***: Both the client and node keep a record of other client's and node's penalty points. Penalty points are used as the basis of a disconnecting mechanism to avoid denial of service (DOS) attacks. When illegal blocks and transactions occur, the originating node will incur penalty points. Then, when the points total reaches 100, all connections from it will be rejected for 24 hours as a punishment.

***New transaction spreading***: as shown in Fig. 2, when the client generates a new transaction, the command 'inv' is sent to the entry node. This informs the node that there is a new transaction with the id 'tx_hash.' The entry node checks the transaction id in its own transaction database, if it exists, the id is disregarded, if not, it will send the command 'getdata' to request the contents of this transaction. The client replies with the command 'tx' as well as the transaction data or replies 'notfound' otherwise. Then the entry node verifies the transaction. If the transaction is not correct, it will return 'reject,' if it is correct, the transaction will be transmitted to its entry nodes.

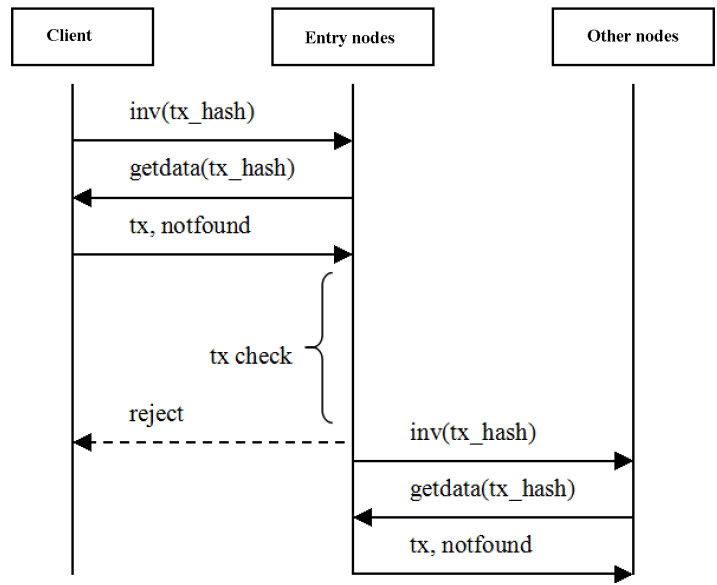

*Fig.2. The process of a new transaction transmitted from client to nodes*

### 2.3. The Sybil attack

In active or passive ABPN, there are several attacks in which the new transaction can be linked to an IP address so that the attacker could find the relationship between the Bitcoin address and IP address.

***Bitcoin Protocol Sniffer***: In the Bitcoin protocol, the data is not encrypted so a well-formed sniffer could monitor all outgoing 'inv' commands to check whether the transaction id has been seen before. If it has then it is likely to be a new transaction. Hence, we could obtain the relationship between the Bitcoin address and the IP address. If such a monitoring system is deployed at the network entrance of a city, the real identity can be found with the help of the telecom operator's IP records.

***Sybil Attack***: Kaminsky [9] proposes the Sybil attack using a Bitcoin client to connect to all nodes in the Bitcoin network. The first source IP address of a new transaction is owned by the original sender. Philip Koshy's [10] experiments show that this method can work, but there are three problems remaining: (1) Bitcoin via TOR hides the true IP address, (2) a large number of clients cannot be connected directly so the sender of the new transaction could not be located properly, (3) the same client owns different sessions, different IP addresses and different networks (anonymous and not anonymous), so it is difficult to link the transactions and IP addresses.

***Sybil attack plus entry nodes***: Alex Biryukov [11] implemented a method that makes all Bitcoin nodes deny connections from TOR exit nodes in 2014. At the same time, they succeeded in detecting the entry nodes of a specified client. With these two tricks, they solved the first two problems of the Sybil attack. Those suspicious IP addresses collected in a Sybil attack include the IP address of the sender, IP addresses of entry nodes, and IP addresses of non-entry nodes. Through the delivery time of every mentioned IP addresses, we could probably then find the source IP address of a new transaction. Their results on a Bitcoin test network show that there is a 60% chance of identifying the source IP of a new transaction successfully using this method.

*Fake nodes attack:* Alex Biryukov [12] developed a TOR middleman attack and 'Address cookies' to solve the third problem for the Sybil attack in 2015 via what could be called a fake Bitcoin nodes attack. This works by firstly, establishing a sufficient number of fake TOR exit nodes (the amount should reach 3% of all TOR network exit nodes) and fake Bitcoin nodes (1,000 to 1,500). These fake nodes behave like normal nodes but they run code from the attackers. Then, address cookies aim to identify a certain client even if it uses different IP addresses, different sessions, and different networks.

## 3. Transaction Remote Release

Anonymous networks such as TOR can encrypt the data of the Bitcoin protocol and hide the real IP address of the client but it cannot prevent the TOR network from being disconnected from the Bitcoin system. They are also vulnerable to the 'middleman' attack and 'address cookies' attack. In fact, with these attacks it is hard to hide the real IP address of the sender of a new transaction. Therefore it is desirable to find ways to resist such attacks.

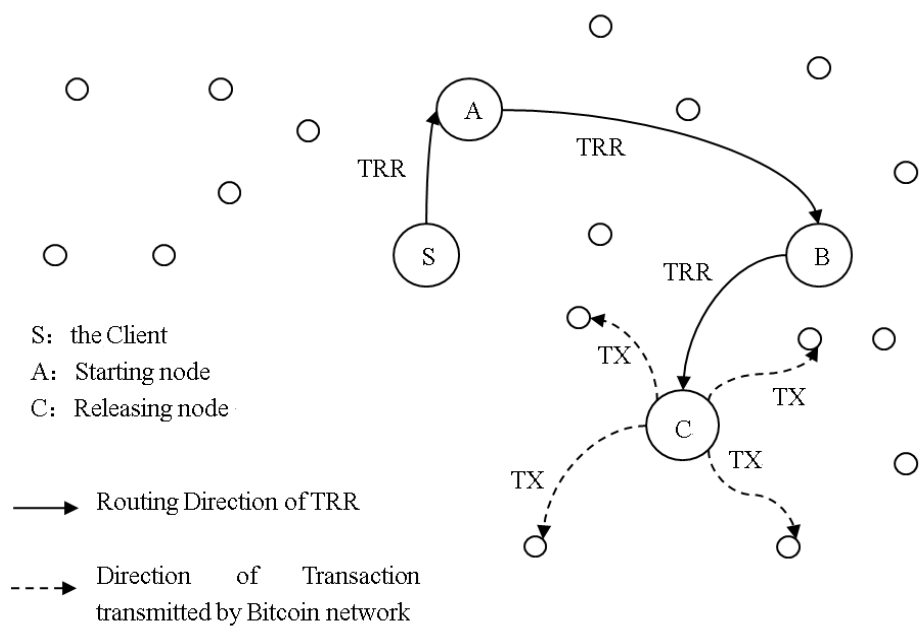

*Fig.3.* The principle of TRR: releasing transactions in remote nodes

In the Bitcoin protocol, the only way that the attackers can connect the Bitcoin address with an IP address is in the process of releasing and spreading a new transaction. If we encrypt the new transaction and obfuscate the source IP of the sender then the attackers may not succeed. Based on the above ideas, this paper puts forward a new anonymization technology called Transaction Remote Release (TRR).

TRR is inspired by the idea of encryption and decryption layer by layer as used in TOR. A client encrypts a new transaction, layer by layer, using the public key from different TRR nodes. Then it establishes an independent connection to other TRR nodes, one by one, without using the spreading mechanism of the Bitcoin network. When a TRR node receives data, it will decrypt it using its private key. Then it transmits the remaining data to the next node. When the last TRR node is reached, it will release the transaction to the Bitcoin network. Every node knows its previous node and next node. Only the client and the last node know the content of the transaction but the last node does not know the IP address of the client.

As shown in Fig. 3, S is the client, while A, B, and C are all Bitcoin nodes on the network. The spreading route is composed of nodes selected randomly by S. TRR data is transmitted to C along the route. C decrypts the transaction then spreads it to the entire Bitcoin network.

### 3.1. TRR algorithm

System setup

(1) Each node has a pair of public keys for data encryption and a private key for decryption. The private key is kept secret.
(2) Each node discloses its public key and IP address to other nodes.
(3) Elliptic Curve Cryptography (ECC) is adopted to encrypt and decrypt data. The selected elliptic curve is secp256k1 [18].
(4) Returned data from TRR nodes is encrypted with the public key generated and provided by the client. The client holds a private key.

In Fig. 3 for example, the TRR algorithm is implemented as follows:
(1) The client S generates a new transaction.
(2) S obtains a list of online TRR nodes and their public keys then selects three to five nodes randomly to form a routing. We call the first node of the routing the 'starting node' and the last node the 'releasing node'. S generates a pair of public keys to encrypt the return data then inserts the public key into the routing information for each node.

(3) S then inserts the new transaction data and releasing delay into the data structure of the TRR request. The request is appended to the routing information used by the corresponding nodes and then it will be encrypted using the public keys from the corresponding node in the opposite order of the routing, one by one. That is, the client S appends the routing information of C to the TRR request and encrypts the entire data with the C's public key. Next, S appends the routing information of B to the encrypted data and encrypts the entire data using the public key of B. Finally, S appends the routing information of A to the encrypted data and encrypts the entire data using A's public key, then sends it to A. The releasing delay refers to the time after which the releasing node should release the transaction and it is counted by the number of Bitcoin blocks. The maximum is 5.

(4) A receives a TRR request and uses its private key to decrypt the data. Then it finds the IP address and port of B in the routing information area and transfers the data to B. B repeats the process and transfers the remaining data to C.

(5) C receives and decrypts the data. If there is no following IP address, it stores the transaction in the transaction memory pool of TRR and then releases it after the releasing delay. C verifies the transaction again before releasing and checks whether this transaction exists in the Bitcoin blockchain or memory pool. If it is found it does nothing. If it is not, C will spread the transaction to the Bitcoin network.

(6) C returns the result when it is successful or returns an error message when it detects an error. When returning data C encrypts the data with the public-key in the routing information and returns the result data back to the client. Once the client receives the data, it will decrypt it, layer by layer then check the reason and then processes it accordingly.

(7) Back to step (2), S continues to send another 1 to 2 TRR requests. The routing of each request is randomly selected and the releasing delay can be changed.

(8) After the releasing delay, S checks whether it has received the transaction or not. If it has, this indicates a successful release. Otherwise the process reverts back to step (2).

*3.2. Analysis of the TRR algorithm*

*3.2.1 Public key management*: public key management consists of generation and broadcasting.

*Generation of the public key pair*: The ECC system is used to generate the public key (the curve selected is secp256k1). The public key is disclosed to other nodes in order to encrypt the data. The private key is kept secret and used to decrypt the data by the node which owns it.

*Broadcast of the public key*: The public key is added to the node and then spread to the entire network. There are two transmission methods: one is to transmit using the command 'addr' within the Bitcoin protocol. The second is through Bitcoin seeders which collect information on all Bitcoin nodes. When clients or nodes start up, they will receive the list of nodes and public keys.

*3.2.2 Routing*: The client selects several nodes from the active nodes in the network at random and arranges them in random order to form a route.

We define dishonest nodes as those which disclose a false public key, deny TRR connections, fail to transmit TRR data or do not release transactions. If there is a node in the selected route which is dishonest, the TRR request will fail.

The success rate of transaction releasing (SRTR) helps us to know whether a route is good or not. Obviously, the higher the SRTR, the more it meets our requirements. Factors that affect it are the number of nodes and the number of routes.

*The number of nodes*: This refers to the number of nodes in a route. The more nodes in the route, the lower the success rate will be in transaction releasing.

*The number of routes:* The SRTR of two or more routes is higher than when relying on a single route.

As an example, suppose that the number of Bitcoin nodes is 6,000 and the number of dishonest nodes is 600. Also, assume there are 1 to 10 nodes in a route and that 1 to 3 routes are selected. It can be seen from Fig. 4 (a) with 2 routes and 2 to 5 nodes, the SRTR is 96.36%, 92.65%, 88.17%, and 83.22% respectively. With 3 routes and 2 to 5 nodes, the releasing rate is 99.31%, 98.01%, 95.93%, and 93.13% respectively. From this, it is apparent that while the proportion of dishonest nodes is 10% and there are 2 to 5 nodes in a route, using 3 routes provides a high SRTR.

It is also apparent from Fig. 4(b) that while there are 3 routes in a TRR request, the SRTR declines rapidly when the dishonesty rate of nodes goes up. The SRTR of 4 to 5 nodes reaches 56.1% and 42.4% respectively when the dishonest rate of nodes reaches 30%. In this case, more than 3 routes are required to raise the SRTR to an acceptable level.

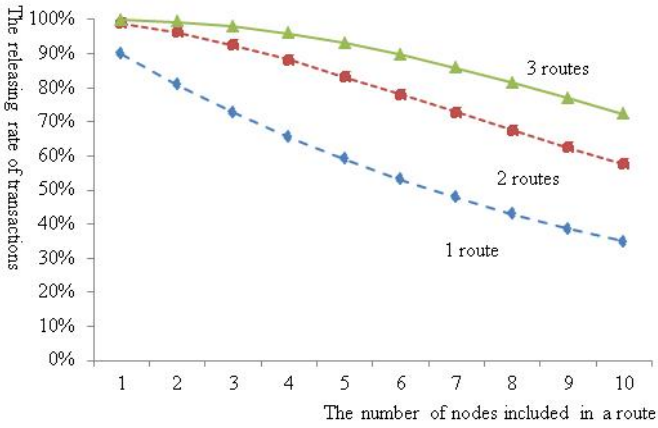 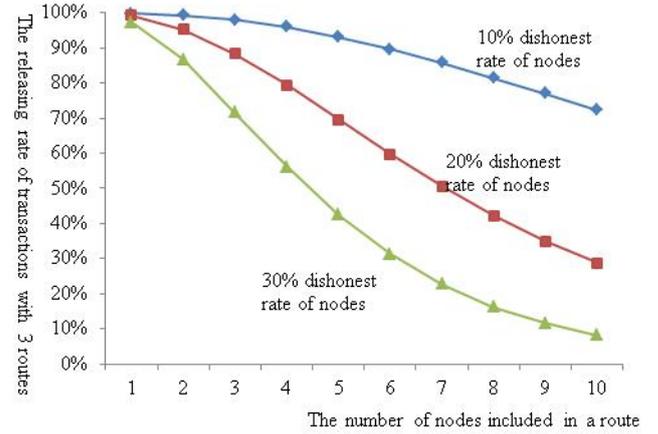

(a) Effect of the number of routes                (b) Effect of the dishonest rate of nodes

Fig.4. The Effect of the number of routes and nodes in a route, the dishonest rate of nodes on the SRTR

*Connection and timeout:* TRR uses short TCP connections and the TRR connections are included in the 117 incoming connections. When TRR data returns or the connection is timed out, the previous node will cut off the connection immediately. When the releasing node returns the processing results, the established connection should shut down node by node.

*3.2.3 TRR protocol*: The client generates transactions and a TRR request. Table 1 shows a data structure named 'trr_data' which comprises a TRR request. This is what the releasing node observes after decryption.

**Table 1** TRR request trr_data

| size | Name | data type | Note |
|---|---|---|---|
| 1 | Version | uint8_t | Version of trr_data |
| 4 | Time | uint32_t | Time |
| 8 | Release time | uint64_t | Releasing delay, it is a block number |
| 2 | tx_size | uint16_t | Size of transaction |
| ? | Tx | byte[] | Transaction content |

The data structure head of the routing named 'trr_routing' is shown in table 2. This is the routing information about the next node received by each node. If there is no next node, the IP address is zero.

**Table 2** TRR routing trr_routing

| size | name | data type | note |
|---|---|---|---|
| 1 | version | uint8_t | Version of trr_data |
| 32 | pubkey | byte[] | Public key used to encrypt returned data |
| 4 | dst_ip | uint32_t | IP address of the next node |
| 2 | port | uint16_t | Port of the next node |
| 2 | tx_size | uint16_t | Encrypted data size |
| tx_size | trr_data | byte[] | transaction |

The returned data 'trr_ack' of each node are shown in Table 3 below:

Table 3 TRR returned data trr_ack

| size | name | data type | note |
|---|---|---|---|
| 1 | version | uint8_t | Version of trr_data |
| 4 | time | uint32_t | Time |
| 4 | rpt_ip | uint32_t | Node reporting this error |
| 4 | err_ip | uint32_t | Node who encounter |
| 2 | errno | uint16_t | Error number, 0 is right |
| 30 | errmsg | char[30] | Error message |

When the command 'trrack' returns data 'trr_ack', it is encrypted by the reporting node using the public key provided by the client. Other nodes transmit the data directly. The client decrypts the data and obtains a 'trr_ack' structure.

TRR simplifies the protocol process because (1) there is no need for a new block refreshing command and a new transaction refreshing command (2) there is no need for address exchange (3) the short connections shut down once the communication has finished (4) the command 'version' is replaced by 'vertrr', which can be used to identify TRR connections; when the command 'vertrr' is received then the connection is a TRR; otherwise a Bitcoin connection is indicated. (4) newly added commands: TRR request 'trr' and TRR ack 'trrack'. The communication process is as shown in Fig. 5.

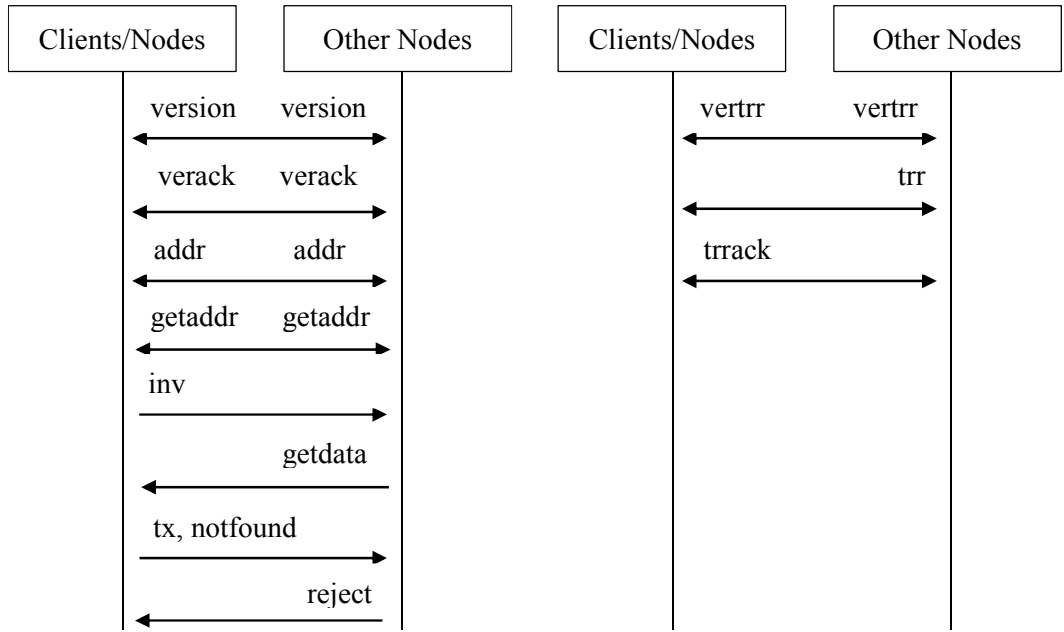

    (a)  Connection in the Bitcoin protocol    (b)  Simplified connection in the TRR protocol

**Fig. 5.** *TRR simplified protocol when performing TRR connections*

*3.2.4 Releasing process*: When C finds that the IP address of the next node is 0, it knows that it is a release node. C verifies the transaction and then puts it into the TRR transaction memory pool (maximum capacity of 1,000). C will then spread the transaction in the usual way as a normal transaction being released after reaching the releasing delay.

We set up a releasing delay of 1 to 5 blocks which is used to disguise the time of each transaction broadcast to the entire Bitcoin network. Thus the attacker could not link the transaction and TRR request by timing sequence. According to the statistics of the Bitcoin blockchain, there are 4,631 blocks and 4,379,983 transactions between block number 344,501 and 344,520. Also, there are on average 945 transactions in every block. If the releasing delay is set to 1, 3, and 5 blocks, then there are 945, 1,890, and 4,725 transactions being mixed into the TRR releasing transaction. Hence, it will be very difficult to deanonymize the transaction through the timing sequence of TRR requests and transaction releasing.

*3.2.5 Attacks against TRR*:

**Bitcoin protocol sniffer**: TRR adopts an asymmetric encryption algorithm to encrypt the new transaction within the Bitcoin protocol. The sniffer can monitor the TRR requests but it cannot decrypt the data.

***Sybil attack***: the attacker connects to all Bitcoin nodes according to the timing sequence of the transactions to determine which transaction came from which client. TRR allows the client to hide the real IP address without TOR and does not directly use the command '*tx*' to transmit transactions. When the releasing node does so later, the timing sequence is delayed from what it would otherwise be. Therefore the attacker cannot find the IP of the sender but only the IP of the releasing node.

***Sybil attack plus entry nodes***: all Bitcoin nodes are connected in the Sybil attack method. The attacker determines whether a transaction belongs to a client or not according to the transmitting speed and transmitted quantity of the entrance nodes. However TRR bypasses the entrance node using a set of randomly selected nodes. Even if the attacker finds the 8 entrance nodes of the client they still cannot find any new transaction launched by the client through those nodes.

***Fake nodes attack:*** The TOR middleman attack uses fake Bitcoin nodes which become the entrance nodes for clients using TOR. 'Address cookies' make unique marks on every client to cluster different sessions, IP addresses and transactions from one client. However TRR makes use of randomly selected routing, bypassing the faked entrance nodes and hiding the real IP address. This makes the attacker regard the releasing node as the sender.

Furthermore, known attacks against the Bitcoin protocol such as Bitcoin protocol sniffer, the Sybil attack, Sybil attack plus entry nodes and fake nodes attack have no significant effect on TRR.

***Fake TRR nodes attack:*** We build on the fake nodes attack to enable it to attack TRR and called it the fake TRR nodes attack. That is, based on the fake nodes attack, some features should be appended such as supporting the TRR protocol, dealing with TRR requests honestly and recording the information of each TRR request, including the IP and port of the current node, the previous node and next node as well as the transaction contents. We call these nodes TRR nodes. These fake TRR nodes collect the above information to a centralized program to recover each route.

We define successful deanonymization as getting the correct route and the contents of the transaction. There are some requirements for a successful fake TRR nodes attack: (1) all fake TRR nodes are honest, because dishonest nodes lead to unsuccessful releasing of transactions and deanonymization fails. (2) The starting node and the releasing node must be TRR nodes and the transaction content and the IP address of the releasing node should be obtained correctly through a fake nodes attack. (3) After one non-fake TRR node in a route, there should be a fake TRR node behind it so that the whole route from the releasing node to the client can be recovered.

Hence, a fake TRR nodes attack depends on the proportion of TRR nodes. Suppose that the number of Bitcoin nodes is 6000, there are 1, 2, and 3 routes, it could be seen from Fig. 6(a) that the success rate of deanonymization (SRD) goes up when the number of routes rises but it declines rapidly when the number of nodes goes up. The SRD reaches no more than 3.0% even if there are 3 routes and there are 2 to 3 nodes in each route; when there are 4 to 5 nodes in each route, the SRD is below 0.6%.

As shown on Fig. 6 (b), we set up 3 levels of the rate of fake TRR nodes (RFTN) including 10%, 20%, and 30%, and then select 3 routes, the SRD and the RFTN are positively correlated. When the RFTN reaches 30% and there are 2 to 5 nodes in each route, the SRDs reach 24.6%, 24.6%, 15.3%, and 6.36% respectively, which are significantly higher than those at the level of 10% of the RFTN. However, as for 4 to 5 nodes, the number of SRDs is too low to be applied widely.

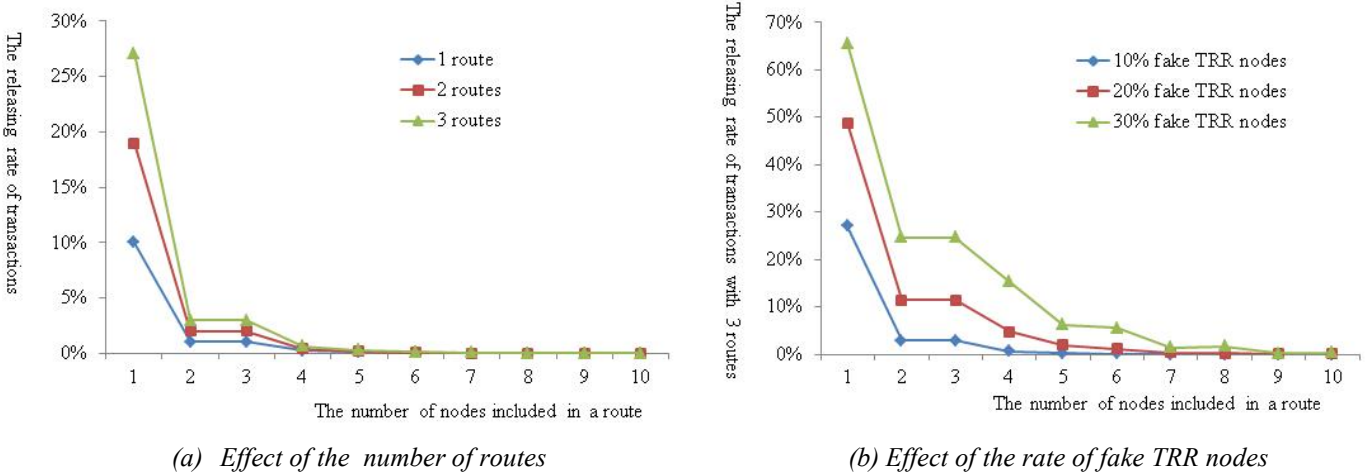

*(a) Effect of the number of routes*     *(b) Effect of the rate of fake TRR nodes*

*Fig.6. The effect of the number of routes and nodes in a route, the rate of fake TRR nodes on the SRD*

## 4. Experiments and Analysis

The implementation of TRR consists of five components: 1) generating and broadcasting public keys; 2) nodes management; 3) route selection; 4) transaction releasing; 5) data encryption and decryption.. Among them, the first 4 items do not present any significant difficulties so they will not be discussed further. However, data encryption and decryption may affect data throughput and response efficiency of the TRR nodes so we will investigate this further by means of a set of experiments.

### 4.1. ECC encryption and decryption algorithm

We will demonstrate the principle of the ECC encryption and decryption algorithm through the following scheme where user A will send encrypted data to user B.

**Step.1** B selects a private key k, and generates a public key $P = kG$, then makes A aware of P.

**Step.2** A encodes a short plain text (a crypto block) to a point M in elliptic curve secp256k1. This is repeated until all plain text is encoded. A then generates a random integer r (r < n) where n is the order of base point G.

**Step.3** A calculates C1 = M+rK, C2 = rG, (C1, C2) which is the cipher text. A sends the cipher to B.

**Step.4** B calculates M = C1-kC2, because C1-kC2 = M+rP-K(rG) = M+rP-r(kG) = M.

**Step.5** B decodes point M, and then gets a short plain text. B repeats calculating M and decoding M, until all the text has been decoded.

Based on the above algorithm and using the open-source math library libtommath [20], we implemented a layer by layer encryption and decryption algorithm based on secp256k1. In addition, we performed an experiment on the length of crypto blocks and another performance experiment on TRR multi-layer encryption and decryption.

### 4.2. The length of the crypto block

The length of crypto blocks will affect the performance of encryption and decryption as well as the growth rate of cipher. Through a combination of experimentation and analysis, we chose 64 bytes as the length of the crypto block. One reason for this is that the length of the private key, which forms the elliptic curve secp256k1 is 256 bits. That is 32 bytes plus the maximum length of the axes of elliptic curve secp256k1 i.e. 32 bytes. Hence, we can encode a point on secp256k1 using 64 bits. Furthermore, the results of our experiments showed that 64 bytes was a reasonable value. We used 16 bytes, 32 bytes and 64 bytes, 3 levels of encryption and made use of 5 public keys from 5 nodes to encrypt 8 groups of plain text with different lengths, one by one. Then we used 5 private keys, one by one, to decrypt the encrypted data, recording the elapsed time and cipher length after every completed run. We also monitored the influence of these 3 length levels on the performance of the encryption and cipher length. The results are shown in Fig. 5. Figure 5(a) shows: that when the length of plain text ranges from 8 bytes to 10,240 bytes, the elapsed time of 5 turns of encryption and decryption at the 64 bytes level is the shortest. Figure 5(b) shows: that when the length of plain text ranges from 8 bytes to 10,240 bytes, the length of cipher generated by a 64 byte crypto block is also the shortest. Furthermore, 64 bytes was earlier found to be a suitable length for the crypto block.

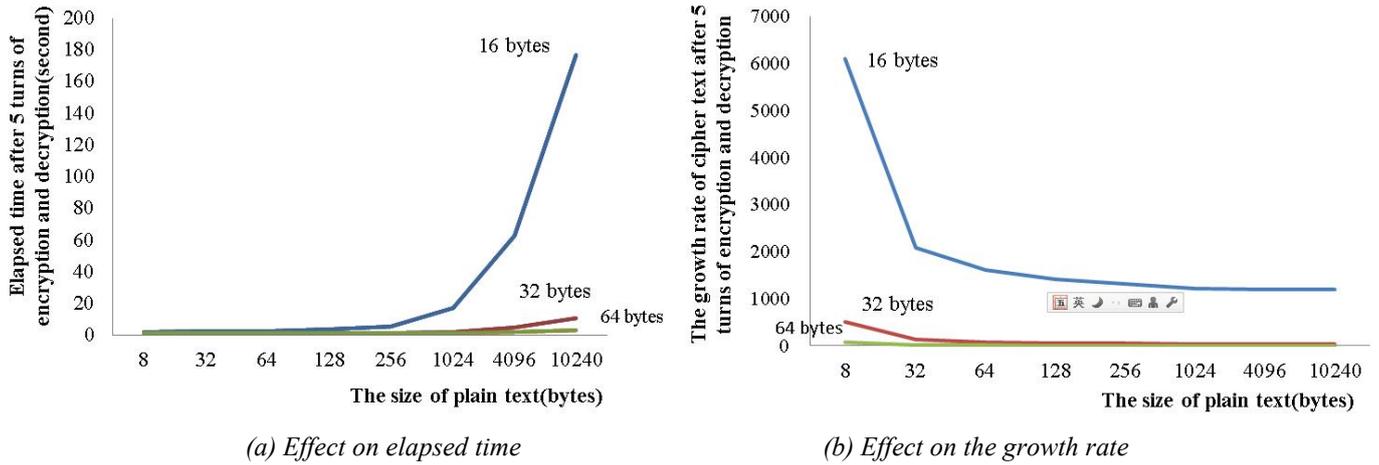

*(a) Effect on elapsed time*    *(b) Effect on the growth rate*

**Fig.5.** *Comparison experiments among 3 different lengths of crypto block*

### 4.3. The algorithm performance of TRR multi-layered encryption and decryption

We now evaluate the algorithm performance of the TRR multi-layer encryption and decryption. Several plain texts with different lengths were encrypted by public keys from 5 nodes, one by one, and then were decrypted by private keys from the same 5 nodes, one by one, using a crypto block of length 64 bytes After every turn, the elapsed time and cipher length were recorded. The results are shown in Table 4. This experiment was run on an Intel Core i7-5500 2.4GHz CPU/16G Memory under Windows 8.

**Table 4** The growth rate of plain text and elapsed time after 5 turns of encryption and decryption

| Plain text size(bytes) | 8 | 32 | 64 | 128 | 256 | 1024 | 4096 | 10240 |
| --- | --- | --- | --- | --- | --- | --- | --- | --- |
| Growth rate (100%) | 82.5 | 20.6 | 10.3 | 5.67 | 3.35 | 1.61 | 1.30 | 1.22 |
| Elapsed time(seconds) | 1.73 | 1.6 | 1.63 | 1.7 | 1.64 | 1.66 | 1.89 | 3.05 |
| Elapsed time of encryption(seconds) | 1.13 | 1.07 | 1.1 | 1.13 | 1.09 | 1.1 | 1.22 | 1.94 |
| Elapsed time of decryption(seconds) | 0.6 | 0.53 | 0.53 | 0.57 | 0.55 | 0.56 | 0.67 | 1.11 |

It can be seen from Table 4, the longer the plaintext is, the slower the cipher text grows. The length of 'trr_data' and 'trr_routing' for two data structures is 46 bytes and the length of a Bitcoin transaction is from 200 bytes to 10,240 bytes. Therefore, the data length of a TRR request is in the range from 256 bytes to 10,240 bytes. Here, the corresponding growth rate of cipher text is from 3.35 to 1.22, that is, the starting node will receive a TRR request whose data length is from 858 bytes to 12,493 bytes. This data growth rate was judged to be acceptable.

In addition, it took the client 1.09 to 1.94 seconds to encrypt plaintext of length 256 bytes to 10,240 bytes using 5 private keys, one by one. As for the client, the time consumed is negligible. Every node spends 0.11 to 0.225 seconds in decrypting the TRR request data, so under the situation of a few TRR requests, it is easy to complete the decrypting operation. According to the statistics of Bitcoin transactions, in July 2015, the average number of transactions daily was 141,290. If these transactions were released through TRR requests, assuming the TRR encryption time is 0.225 seconds, then every node could decrypt all TRR requests within 8.8 hours. Hence, the algorithm of TRR multi-layered encryption and decryption is feasible in practice.

*4.4. Analysis of the advantages and disadvantages of TRR*

TRR borrows the idea of routing and multi-layered encryption from TOR but it is a new way of transmitting Bitcoin transactions. A comparison between TOR and Bitcoin is shown in Table 5.

Table 5 Comparison of the transaction spreading method among TRR, TOR and the Bitcoin network

| Items | | TRR | TOR | Bitcoin network |
|---|---|---|---|---|
| Data encrypted | | New transactions | All the data | No encryption |
| Stability of connections | | Stable | Not Stable | Stable |
| Speed of connections | | Faster | Slow, blocked sometime | Connections exist, very fast |
| Exit nodes | | None | Yes, vulnerable to attacks | None |
| Data transfer | | Certain direction | Certain direction | Broadcast in entire network |
| Anonymity Effect | Protocol sniffer | No | No | Yes |
| | Sybil attack | No | No | Yes |
| | Sybil + Entry nodes | No | Yes(TOR Cut off) | Yes |
| | Fake Bitcoin nodes | No | Yes (TOR Middle-man) | Yes |
| | Anonymity | Strong | Normal | Weak |
| Implement Difficulty | | Hard, to modify Bitcoin protocol | Supported | Supported |

The TOR protocol may transmit any kind of data; however TRR only can be used for transmitting Bitcoin transactions. Compared to TOR, TRR has the following advantages.

(1) Encrypting new transactions is an effective method of preventing attackers from clustering Bitcoin addresses and the IP address of the sender. Although the Bitcoin blockchain is public and does not need to be kept secret, TOR encrypts all blockchain data unnecessarily but TRR only encrypts and transmits new transactions, thus improving the performance and throughput of nodes.

(2) Improving the communication performance. TOR has been prohibited or blocked by many countries while TRR has not been affected because it will not be in conflict with these countries' policy. Hence the connection of TRR could be more stable and faster.

(3) Less vulnerability to a middleman attack. TOR exit nodes are vulnerable to cut off or middleman attacks. Fortunately, there are no specific exit nodes in the Bitcoin network, any Bitcoin node may become the focus of connections and thus TRR will be immune to cut off or middle-man attacks.

(4) The results of our experiments show that the performance of the TRR multi-layered encryption and decryption algorithm is good and the loss of clients and nodes is very small. Meanwhile the growth of the cipher caused by encryption is very limited.

However, TRR has two weaknesses:

(1) It would be necessary to modify the Bitcoin protocol in order to apply TRR technology but TOR is well established in serving the Bitcoin network.

(2) There is no way currently to determine whether a client is honest or not so TRR is vulnerable to fake TRR requests and Denial of Service (DOS) attacks based on it.

## 5. Conclusions

This paper analyses deanonymization attacks against the Bitcoin Protocol and Network, such as Bitcoin protocol sniffer, the Sybil attack, Sybil attack plus entry nodes, fake Bitcoin nodes and fake TRR nodes attack. The paper proposes a new anonymization technology named Transaction Remote Release (TRR), and it describes the principles and anonymization in TRR then discusses possible attacks against it. The analysis results show:

(1) When 3 routes are selected and there are 5 nodes in each route, the SRTR reaches 93% when the dishonest rate of nodes is 10% and the SRD hits no more than 0.6% when the RFTN is 10%.

(2) TRR can prevent several known attacks against the Bitcoin Protocol and Network to maintain anonymity. We constructed a fake TRR nodes attack and the analysis results show that with 3 routes selected and 5 nodes in each route, the SRD is less than 6.3% even if the RFTN reaches 30%. Hence, TRR can help clients gain strong anonymity.

(3) In addition, the experiments show that the performance of the TRR multi-layered encryption and decryption algorithm is satisfactory in practice and the growth rate of cipher text is very limited.

The weakness of TRR is the need to modify the Bitcoin protocol; it is also vulnerable to DOS attacks based on fake TRR requests, and this is the direction of further research.

For faster implementation and application of TRR technology, setting up a new blockchain or establishing a Transaction Delivery Network (TDN) are two possible ideas avenues for further research. DarkNetspace is an independent blockchain of crypto-currency, based on TRR technology to enhance the anonymity of currency transactions. TDN is an independent network based on TRR technology to distribute new transactions from any blockchain anonymously, supporting multi-currencies and multi-blockchains.